# Dimensional Crossover and Topological Phase Transition in Dirac Semimetal Na$_3$Bi Films

Huinan Xia[1,#], Yang Li[2,3,4,#], Min Cai[1], Le Qin[1], Nianlong Zou[2,3,4], Lang Peng[1], Wenhui Duan[2,3,5], Yong Xu[2,3,4,†], Wenhao Zhang[1,*], Ying-Shuang Fu[1,‡]

1. *School of Physics and Wuhan National High Magnetic Field Center, Huazhong University of Science and Technology, Wuhan 430074, China*

2. *State Key Laboratory of Low-Dimensional Quantum Physics, Department of Physics, Tsinghua University, Beijing 100084, China*

3. *Collaborative Innovation Center of Quantum Matter, Tsinghua University, Beijing 100084, China*

4. *RIKEN Center for Emergent Matter Science (CEMS), Wako, Saitama 351-0198, Japan*

5. *Institute for Advanced Study, Tsinghua University, Beijing 100084, China*

   [#]These authors contributed equally to this work.

   Email: [*]wenhaozhang@hust.edu.cn, [†]yongxu@mail.tsinghua.edu.cn, [‡]yfu@hust.edu.cn

**ABSTRACT:**

**Three-dimensional (3D) topological Dirac semimetal, when thinned down to 2D few layers, is expected to possess gapped Dirac nodes *via* quantum confinement effect and concomitantly display the intriguing quantum spin Hall (QSH) insulator phase. However, the 3D-to-2D crossover and the associated topological phase transition, which is valuable for understanding the topological quantum phases, remain unexplored. Here, we synthesize high-quality Na$_3$Bi thin films with √3×√3 reconstruction on graphene, and systematically characterize their thickness-dependent electronic and topological properties by scanning tunneling**

microscopy/spectroscopy in combination with first-principles calculations. We demonstrate that Dirac gaps emerge in Na$_3$Bi films, providing spectroscopic evidences of dimensional crossover from a 3D semimetal to a 2D topological insulator. Importantly, the Dirac gaps are revealed to be of sizable magnitudes on 3 and 4 monolayers (72 and 65 meV, respectively) with topologically nontrivial edge states. Moreover, the Fermi energy of a Na$_3$Bi film can be tuned *via* certain growth process, thus offering a viable way for achieving charge neutrality in transport. The feasibility of controlling Dirac gap opening and charge neutrality enables realizing intrinsic high-temperature QSH effect in Na$_3$Bi films and achieving potential applications in topological devices.

Topological Dirac semimetals (DSMs) host 3D Dirac fermions that disperse linearly along all three momentum directions around the Dirac points,[1,2] distinctive from their 2D analog in graphene or on the surfaces of topological insulators (TIs). Candidate materials, including $Cd_3As_2$ and $Na_3Bi$, have been theoretically predicted[3,4] and experimentally verified to be 3D topological DSMs.[5-10] Due to their quantum criticality neighboring variant topological phases, topological DSMs can be tuned into exotic phases of Weyl semimetal, topological superconductor or axion insulator[11,12] by introducing different kinds of symmetry-breaking perturbations.[13] For instance, $Na_3Bi_{1-x}Sb_x$ and $Cd_3[As_{1-x}P_x]_2$ undergo a topological DSM to trivial insulator transition upon concentration alloying.[14] Black phosphorus and α-Sn can be driven into DSM phase with electric field from surface potassium doping[15] and uniaxial tensile strain,[16,17] respectively. These progresses make topological DSMs an ideal platform for the realization of exotic topological quantum physics and device applications.

Dimensionality provides an additional degree of freedom that can bring about fruitful phenomena to the topological DSMs. When the film thickness of DSM is comparable to the quasiparticle wavelength, finite-size effect quantizes the wave vector perpendicular to the film plane, resulting in 2D massive Dirac fermions in the bulk.[3,4,18] Depending on specific surface orientation and thickness of the films, distinct topological quantum phases may be generated. For film surface quantized along the connecting direction of the two Dirac nodes of the DSM, as exemplified in $Cd_3As_2$(112) films, their topological Fermi arc surface states are kept gapless, which host 3D quantum Hall effect from Weyl orbits.[19-21] On the other hand, for film surface quantized perpendicular to the connection of the two Dirac nodes, for instance in $Na_3Bi$(001) films, the Dirac gapped bulk states are theoretically predicted to convert between QSH and trivial insulators with varying film thickness in an oscillatory fashion.[3,4,22] Specific theoretical predictions are, however controversial, which either predicts a trivial insulator below 7 monolayers (MLs)[23] or a dual 2D TI and topological crystalline insulator in the monolayer.[24] In contrast to $Cd_3As_2$(112) films, rare predictions are experimentally realized for quantum confined $Na_3Bi$(001) films.[25] Until very recently, monolayer and double layer $Na_3Bi$ films on Si(111) substrate are found to be TIs with sizable bulk gap opening.[26] However, the 3D-to-2D dimensional crossover, a central issue to the topological physics, and the associated topological phase transition in 2D Dirac gapped states are still left unexplored.

In this study, we grow high-quality $Na_3Bi$ films with a √3×√3 surface reconstruction of variant thicknesses by molecular beam epitaxy on a graphene-covered 6H-SiC(0001) substrate, which has weak van der Waals interactions with $Na_3Bi$ films. Through proper growth condition, we are able to *in situ* tune the Fermi energy ($E_F$) of $Na_3Bi$ film and thus achieve charge neutrality point of carriers. By scanning tunneling microscopy/spectroscopy (STM/STS) and density functional theory (DFT) calculations, we clarify that $Na_3Bi$ experiences a 3D-to-2D transition from a bulk semimetal to a gapped insulator in form of ultrathin films, where a Dirac band gap of ~72 meV (~65 meV) is observed for 3 MLs (4 MLs) due to quantum confinement. Quasiparticle interference imaging reveals the characteristic Dirac-cone states in both gapped and gapless $Na_3Bi$ films. Furthermore, our observations of edge states, whose energies locate inside the insulating bulk gap and have a decay length comparable with our DFT calculations, support their ascription as topological edge states. Our study illustrates the dimensional crossover and topological phase transition of $Na_3Bi$(001) films, and demonstrates its viability for carrier control, which are all valuable for realization of QSH transport and device applications.

**RESULTS AND DISCUSSION**

Bulk Na$_3$Bi crystallizes in a hexagonal P6$_3$/*mmc* phase, whose ML is composed of a Na-(Na/Bi)-Na sandwiched structure with an in-plane lattice constant of 5.448 Å.[3,5] Figure 1a shows a typical STM topographic image of as-grown Na$_3$Bi film on a graphene substrate, containing various thicknesses of triangle or hexagonal islands. The zoom-in image in Figure 1b shows the atomic resolution of Na$_3$Bi surface. Its fast Fourier transformation (FFT) (Figure 1b inset) resolves two groups of spots (red and yellow circles), which are relatively rotated by 30°. They correspond to two real-space lattices of 5.5 Å and 9.5 Å, stemming from the (1×1) Na-terminated atomic lattice[27] and a √3×√3 surface reconstruction, respectively. This surface reconstruction is robust on all Na$_3$Bi films, irrespective of the film thickness or growth condition (see the Supporting Information Figure S1), excluding the electronic origin of such a surface termination. Although similar √3×√3 reconstructions have been reported previously on a Na$_3$Bi/Si(111) film by STM,[28] the detailed structure associated with the influence on its topological properties remain unknown.

Based on DFT calculations and *ab initio* random structure searching (Method section), we have considered thousands possible structures of freestanding Na$_3$Bi films and identified the most stable configuration of the √3×√3 reconstruction. As displayed in Figure 1c for monolayer Na$_3$Bi, the basic bonding framework and chemical stoichiometry do not change upon reconstruction. However, a trimer is formed by three neighboring Na atoms in the upper sublayer, but no trimer is formed in the bottom sublayer due to inversion asymmetry. Adjacent Na atoms rearrange away from the trimers so as to reduce Pauli and Coulomb repulsions between Na atoms, which leads to a structural buckling of ~1.55 Å in the middle sublayer. Consequently, the mirror symmetry in the out-of-plane direction ($M_z$) gets broken, while the $C_{3v}$ point group symmetry is preserved. Similar √3×√3 reconstructions are found for thicker films (Figure S2), in consistent with experiments. By comparing band structures of Na$_3$Bi thin films before and after the reconstruction (Figure S3), we conclude that the reconstruction has minor influences on band structure, despite noticeable Rashba-type spin splitting in the monolayer caused by M$_z$ symmetry breaking. This is in accordance with the folding bands of Fermi surface features between the K and Γ points in previous angle-resolved photoemission spectroscopy (ARPES) experiments.[29]

A considerable amount of defects are randomly distributed on the as-grown surface (Figure 1b). The typical d$I$/d$V$ spectra show a distinct minimal dip (blue arrow in Figure 1d) at ~188 meV, and the conductance monotonically increases on both sides of the dip. This is in line with the suppressed local density of states expected from the bulk Dirac states of Na$_3$Bi, where the spectroscopic dip corresponds to the Dirac point $E_D$.[30,31] The spectra measured along a straight line crossing the abundant defects are rather homogeneous. Probably, the high defect density increases doping carriers ($E_D$ ~188 meV), which may reduce the screening length.[30] The $E_D$ above $E_F$ indicates the as-grown sample with dense Na vacancies is heavily *p*-doped and the dominated carriers are hole-like. By increasing Na flux during growth, we find the surface prevails with the ordered √3×√3 reconstruction of sparse impurities (Figure 1e). The corresponding STS (blue curve in Figure 1f) indicates the sample changes into *n*-type doped with $E_D$ located at -73 meV. This transition from hole- to electron-like doping is reasonable since Na vacancies usually act as charge acceptors. Certainly, we post-anneal the as-grown Na$_3$Bi film in an extra Na flux, which shifts its $E_D$ to -5 meV (red curve in Figure 1f), approaching the charge neutrality point. Conversely, for a pristine *n*-type Na$_3$Bi film grown under large Na flux, we can also convert it to *p*-doping by increasing Bi flux or by annealing in vacuum without Na flux (Figure S4). In contrast to previous methods of carrier tuning of Na$_3$Bi by surface molecular or electrostatic gating,[7,31] our approach of *in situ* growth adjustment extends the carrier control from the surface to the bulk states, which is more uniform throughout the film.

Next, we investigate the electronic structure of Na$_3$Bi film as a function of thickness from d$I$/d$V$ spectra (Figure 2a). For the 3-ML Na$_3$Bi film, the thinnest island we obtain, the d$I$/d$V$ becomes vanished near $E_F$, showing a clear gap of ~72 meV. The gap becomes slightly smaller on the 4 MLs (~65 meV). However, for the film thickness >5 MLs, finite conductance persists in all energies. There exists a spectroscopic dip around $E_D$ for 5-8 MLs, reminiscent of a band gap. The dip becomes less clear with increasing thickness and completely disappear for thick films, gradually evolving into the bulk limit (Figure S5).[30] These spectroscopic characteristics reveal that the DSM Na$_3$Bi experiences a smooth 3D-to-2D dimensional crossover with gap-opening in reduced dimension, from a bulk gapless DSM to a gapped insulator. Similar phenomena of Dirac gap-opening by varying film thickness have been reported for 3D TI Bi$_2$Se$_3$ films.[22,32]

We further image the band dispersions of the Na$_3$Bi films with the gap opening *via* quasi-particle scattering. 2D conductance plots (Figure 2c,g) are obtained near the step edges of 3- and 5-ML Na$_3$Bi films along two straight lines (Figure 2b,f). Selected d$I$/d$V$ curves can be found in Figure S6. Clearly, spatially modulated standing-wave patterns are formed starting from positive bias voltage (Figure 2c,g), which originate from scatterings by the step edges.[8,9,27] Upon decreasing the energy, the patterns become bent with the diverging wavelength towards the gap edges. The lack of clearly dispersing features in the valence band is probably hindered by the deviation from the linear dispersion, as well as complicated hole bands that lie closely to involve the quasiparticle interference process, which is absent in the conduction band. Analyzing their 1D FFT in Figure 2d,h, we observe linear dispersed branches on both surfaces, coinciding with the scattering modes of Dirac bands of Na$_3$Bi.[5,27] While the 5 MLs film shows a spectroscopic dip around $E_D \sim +10$ meV (Figure 2h), an obvious band gap is opened centered at ~ -100 mV on the 3 MLs film (Figure 2d), conformable with their point spectra (Figure 2c,e and S6). By linear fitting the slope of the dispersion, we obtain the Fermi velocity of 2.20 eV·Å (or 3.34×10$^5$ m·s$^{-1}$) for 3 MLs and 1.94 eV·Å (or 2.95×10$^5$ m·s$^{-1}$) for 5 MLs, respectively, in the same order of magnitude reported by ARPES, STM and transport measurements.[5,27,31] The Dirac-cone dispersion can be also resolved by performing d$I$/d$V$ mapping over a 6-ML hexagonal island at various energies (Figure S7). Notable discrepancy on gap size between theory and experiment exists. This can be understood from a schematic shown in the inset of Figure 2h. The theoretical calculations don't account finite lifetime of the bands (shaded cyan lines), which is contributed by interactions such as electron-electron, electron-phonon, and extrinsic effects such as background tunneling and disorder.[33] Thus, the experimental gap size ($\Delta_{STS}$) gives a lower bound of the theoretical one ($\Delta_{band}$).

To examine the topological properties of the gapped film, we scrutinize the spatial distribution of STS spectra near the step edge of a 3-ML Na$_3$Bi island (Figure 3a). As shown in Figure 3b, the spectrum gradually gains weight inside the band gap upon approaching the step edge, appearing edge states.[26,34] Such localized edge states can be more evidently seen by comparing the spectra taken far away from the edge (black curve) and at the edge (red curve). By integrating the d$I$/d$V$ intensities within the band gap from -60 meV to +20 meV, we trace the distance dependence of the edge state signal in Figure 3c. It exhibits a monotonous decay away from the step edge, whose decay length (λ) is estimated as ~4.1 nm from an exponential fitting. Despite it is weak and notably inhomogeneous, the intensity signal of the edge states is clearly visible, showing explicit 1D feature with approximately uniform width surrounding the periphery of the island (Figure S8). The edge states seem to be more consistent with a topological origin from following aspects: (i) They emerge within the 2D bulk band gap.[26,34] (ii) their decay length is comparable to previously identified topological edge states in a 2D TI of single-layer Bi (111) and 1T′-WTe$_2$,[35,36] instead of trivial ones that are localized within few atomic lattices. (iii) They survive irrespective of the edge inhomogeneity.[37]

To elucidate the experimental observations, we theoretically studied the 3D-to-2D crossover of the topological DSM by performing DFT calculations on electronic and topological properties of Na$_3$Bi films. The influence of substrate and reconstruction is not important as discussed above and thus neglected for simplicity. In monolayer Na$_3$Bi, the major electronic properties are determined by the Na-Bi honeycomb lattice, which, in contrast to graphene, gives large band gaps near the K and K' points caused by the asymmetric sublattices. The low energy physics is dictated by electronic states near Γ. When excluding the spin-orbit coupling (SOC), the valance band maximum is mainly composed of degenerate $p_{xy}$ orbitals of Bi, while the conduction band minimum is mainly contributed by $s/p_z$ orbitals of Bi and Na (Figure 4a). The inclusion of SOC induces an extraordinarily large band spitting (0.85 eV) for the $p_{xy}$ orbitals due to the heavy element and thus leads to an $s$-$p$ band inversion. This SOC-induced band inversion is topologically nontrivial, as confirmed by computation of topological invariant $Z_2$ and edge states, in agreement with previous studies.[23,26] The underlying mechanism is the same as found for other 2D TIs like stanene.[26,34,38]

For Na$_3$Bi films, the interlayer coupling plays a crucial role in defining thickness-dependent material properties, whose influence is significantly stronger for the $s/p_z$ orbitals than for the in-plane $p_{xy}$ orbitals, as evidenced by much larger band splittings of the former orbitals (Figure 4a).

Thus for 2 MLs, an *s-p* band inversion already occurs in the absence of SOC, which is driven by the orbital-dependent interlayer coupling. Without SOC, the system would be a 2D Dirac semimetal protected by the $C_{3v}$ symmetry. The SOC opens a Dirac gap, giving a QSH insulator phase. The same scenario also works for films from 3 MLs (Figure 4b,c) to 7 MLs. The band gap decreases monotonously with increasing thickness, except an abrupt change near 7 MLs. Detailed analysis finds that the interlayer coupling induces an additional *s-p* band inversion above 7 MLs, inducing a transition to a topologically trivial phase. Previous calculations predicted an opposite trend,[3,23] which, however, were done by using the ***k·p*** method that is inadequate to describe ultrathin films. Detailed results of band structure and edge states are presented in Figure S9 and S10. A summary of band gap and $Z_2$ values is presented in Figure 4d. The calculated gap is 46 (27) meV for 3 (4) MLs, slightly smaller than the experimental values. For thicker layers, small band gaps are predicted theoretically but no absolute band gap is observed experimentally. The discrepancy might be caused by the generalized gradient approximation applied in DFT calculations, which typically suffers from a band-gap problem. Moreover, the predicted mini gaps in thicker films, if truly existing, might not be detectable by possibly due to the finite spectroscopic resolution of ~3 meV (see the Method section), spectral broadening from interactions which is not accounted in the calculation. Furthermore, the mini gap may also be influenced by other factors such as the effect of tunneling background, doping, disorder, strain, *etc*.[14,24,39-41] Scrutiny over the spectra of thin films from 5-8 MLs (Figure 2a) indicates the spectral dip around $E_D$, which estimates the gap size on the order of 10 meV from their half-width. We have also estimated the effect of lateral confinement on the gap size, and found its influence is minor (Supplementary Note and Figure S11).

We also present the calculated topological edge states of 3-ML $Na_3Bi$ in Figure 4e, which shows a Dirac-like linear dispersion as expected. Typically, the penetration depth of edge states, which characterizes their real-space distribution, is smallest at the Dirac point and increases to infinite when the edge bands merge into bulk bands (Figure 4e). Here the Dirac point is buried into the bulk valence bands, due to a topological band inversion between the lowest conduction band and the third highest valence band. Thus topological edge states within the bulk gap have a very large $\lambda$ (about 2-3 nm), and are not localized on edge atoms anymore but show a broad distribution. The magnitude of $\lambda$ is consistent with the experimental value for in-gap edge states (Figure 3c, ~4.1 nm), supporting their topological origin. Note that for edge modes with increasingly large $\lambda$, the edge-edge interactions become important when $\lambda$ approaches the size of samples, which could open a hybridization band gap on the edges and thus eliminate the gapless feature.[14] Considering that $\lambda$ of QSH edge states depend sensitively on Fermi energy, the feasibility of tuning $E_F$ demonstrate in our experiments enables controlling the strength of edge-edge interactions, which could be applied to manipulate edge-state transport and is potentially useful for device applications.

**CONCLUSIONS**

To conclude, by combining STM/STS and DFT calculations, we have investigated the thicknesses-dependent electronic properties of $Na_3Bi$ films with simultaneous spatial, energy and momentum resolution. We observe clear signatures that $Na_3Bi$ undergoes a gap-opening transition in dimension-reduced films, a QSH state with a large energy gap and expected edge states that differs from the bulk counterpart. The $E_F$ in $Na_3Bi$ films can be successfully tuned controllably *via* proper growth condition, achieving charge neutrality. Our work provides a route toward studying exotic topological phenomena,[42-45] including realization of quantum spin Hall transport,[46] manipulations of coupling between topological edge states,[47] topological quantum phase transition

with reduced dimensionality,[26] and sets a foundation for topological device applications.[48]

**METHODS**

**Sample Preparation.** To prepare a uniform graphene substrate, a commercial 6H-SiC(0001) wafer, was firstly degassed at 600°C for 3 hours, and then annealed at 950°C under a Si flux for 5 cycles. The SiC substrate became atomically flat terminated with graphene by flashing to 1400°C for 10 minutes. High-purity of Na (99.95%) and Bi (99.999%) were simultaneously co-evaporated from two homemade thermal effusion sources onto the substrate, whose temperature was held at 200°C. During the growth, the Bi:Na flux ratio, controlled by the temperatures of Bi (380°C) and Na (245°C) sources, was kept larger than 1:10. There are two ways to achieve the *n*-doped $Na_3Bi$ films: by raising the Na source to 260°C with a larger flux or by annealing the sample at the growth temperature for 10 minutes in a Na overflux.

**STM/STS characterization.** STM/STS measurements were performed on a Unisoku STM system operating at 4.5 K.[35,37] A W tip with electrochemical etching was cleaned by e-beam heating and calibrated on Ag islands before all measurements. All topographic images were taken in a constant-current mode, and the tunneling d*I*/d*V* spectra and conductance mappings were acquired by standard lock-in technique at 983 Hz with modulation voltage amplitude of 1% of the setting bias voltage. Accordingly, the energy resolution in STS is determined by the thermal broadening and lock-in modulation effects, and is given by: $\Delta E = \sqrt{(2V_{mod,RMS})^2 + (3.2k_BT)^2}$, where $V_{mod,RMS}$ is the root-mean-square value of the modulation voltage amplitude and T is the working temperature (4.5 K).[49] Thus, we obtain the energy resolution of ~ 3.0 mV, which gives an uncertainty of band gap as 6 meV (twice the energy resolution for two gap edges) for thickness >5 MLs.

**First-principles calculations.** *Ab initio* random structure searching (AIRSS)[50] was applied to search stable structures of Na$_3$Bi films. The geometry optimization, total energy, and electronic structure calculations were performed by density functional theory using projector augmented wave potential[51] and the Perdew-Burke-Erzernhof[52] exchange-correlation functional as implemented in the Vienna *ab initio* simulation package.[53] The plane wave basis set with an energy cutoff of 200 eV was used. The slab model with a vacuum layer of 12 Å was applied to simulate thin films. Structure relaxations were carried out until atomic forces is converged to smaller than 0.01 eV/Å. The Brillion-zone was sampled by a 24×24×1 **k**-grid. *Wannier90* and *WannierTools* codes[55,55] were applied to calculate topological invariant and topological edge states.

The Bernevig-Hughes-Zhang (BHZ) model was used to compute the penetration depth of 3-ML Na$_3$Bi as done previously.[56] Band structure of BHZ model is given by: $E_\pm = C - D(k_x^2 + k_y^2) \pm \sqrt{A^2(k_x^2 + k_y^2) + (M - B(k_x^2 + k_y^2))^2}$, where the parameters fitted from DFT-calculated bands near the Γ point are $A$ = 2.58 eV·Å, $B$ = 6.76 eV·Å$^2$, $C$ = -0.165 eV, $D$ = -3.74 eV·Å$^2$, $M$ = 0.220 eV. Dispersion of edge states is given by: $E_{edge} = C + (-DM \pm A\sqrt{B_+ B_-} k)/B$, where $B_\pm = B \pm D$. Based on the above parameters, the group velocity of edge states for 3-ML Na$_3$Bi calculated by the BHZ model is 2.151 eV·Å, which is close to the theoretical value (2.712 eV·Å) obtained by Wannier-function methods and the experimental value (2.20 eV·Å) measured by quasi-particle interference. The reliability of the model and parameters is thus justified. The inverse of penetration depth is given by: $\lambda^{-1} = N - \sqrt{N^2 + (k - k^+)(k - k^-)}$, where $k_\pm = \frac{DN}{B}\left(1 \pm \sqrt{1 + \frac{BM}{D^2 N^2}}\right)$ and $N = A/(2\sqrt{B_+ B_-})$.

**Supporting Information**

The Supporting Information is available free of charge on the ACS Publications website at DOI: xxx.

Figures of bias-dependent STM topographic images with √3×√3 surface reconstruction; figures of calculated structural model for √3-reconstructed $Na_3Bi$ films; figures of influence of √3×√3 surface reconstruction on electronic band structures; figures of tuning $E_D$ by proper growth conditions; figures of d$I$/d$V$ spectra recorded on $Na_3Bi$ terraces with different thickness; figures of selected d$I$/d$V$ spectra for different locations obtained on 3 and 5 MLs; figures of d$I$/d$V$ mappings over a 6-ML hexagonal island at various energies; figures of bias-dependent edge state mappings in real-space near a 3-ML step; figures of calculated band structures for $Na_3Bi$ films with various thicknesses; figures of edge-state calculations for $Na_3Bi$ films with various thicknesses. supplementary note and figure on the lateral confinement effect.

**ACKNOWLEDGMENTS**

We thank Shun-Qing Shen and Gang Xu for discussions. This work is financially supported by the National Key Research and Development Program of China (Grants No. 2017YFA0403501, No. 2018YFA0307000, No. 2016YFA0401003, No. 2016YFA0301001, No. 2018YFA0307100 and No. 2018YFA0305603) and the National Science Foundation of China (Grants No. 11774105, No. 11504056, No. 11522431, No. 11474112, No. 11874161, No. 51788104, No. 11874035, No. 11674188 and No. 11334006) and the Beijing Advanced Innovation Center for Future Chip (ICFC).

# Figures

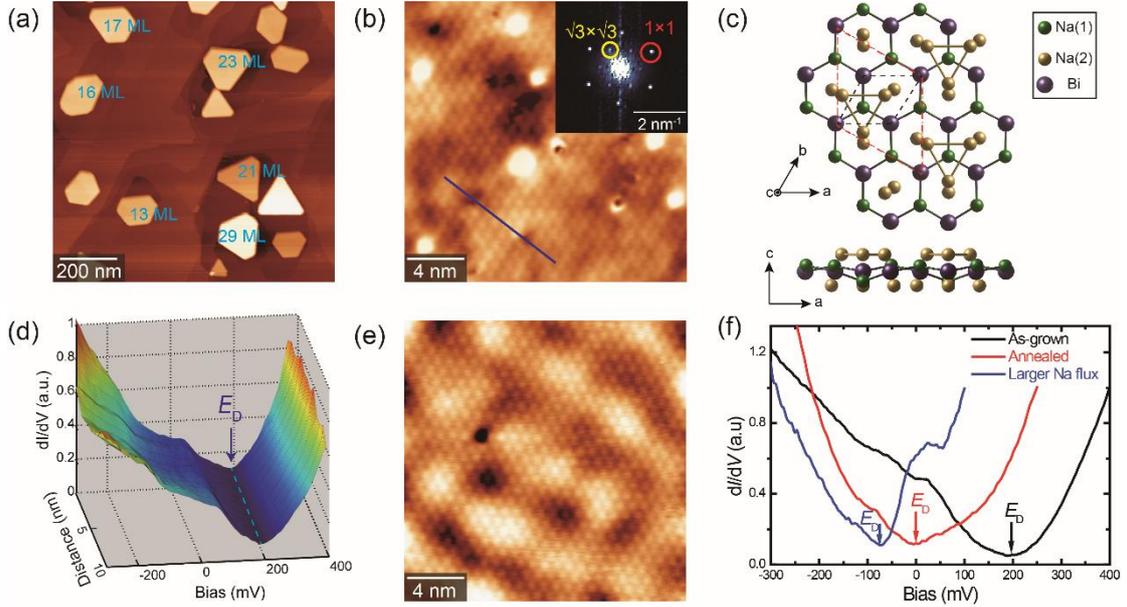

**Figure 1. (a)** STM morphology of Na$_3$Bi film grown on a graphene substrate, with different thicknesses of triangle or hexagonal islands. Scanning condition: $V_{bias}$ = +3.0 V, $I_t$ = 10 pA. **(b)** The atomic-resolution STM image with a periodic lattice of 9.5 Å on the as-grown Na$_3$Bi films ($V_{bias}$ = +100 mV, $I_t$ = 100 pA). Inset is the FFT of (b), where red and yellow circles correspond to the 1×1 Na-terminated lattice and the √3×√3 surface reconstruction, respectively. **(c)** The top and side views of calculated structural model for a most stable Na$_3$Bi film of 1 ML. The 1×1 and √3×√3 reconstruction are marked as black and red dashed rhombus, respectively. **(d)** Spatial variation of d$I$/d$V$ spectra taken along the blue line of 10 nm in (b). The blue arrow marks the intensity minimum at $E_D$. **(e)** The atomic-resolution STM image on a Na$_3$Bi film grown by increasing Na flux ($V_{bias}$ = +100 mV, $I_t$ = 100 pA). **(f)** Comparison for the d$I$/d$V$ spectra taken on different samples: the as-grown Na$_3$Bi films with a small (black) and large (blue) Na flux, as well as Na$_3$Bi films annealed with extra Na flux (red). The corresponding arrows are the positions of $E_D$ as 188 meV, -73 meV and -5 meV, respectively.

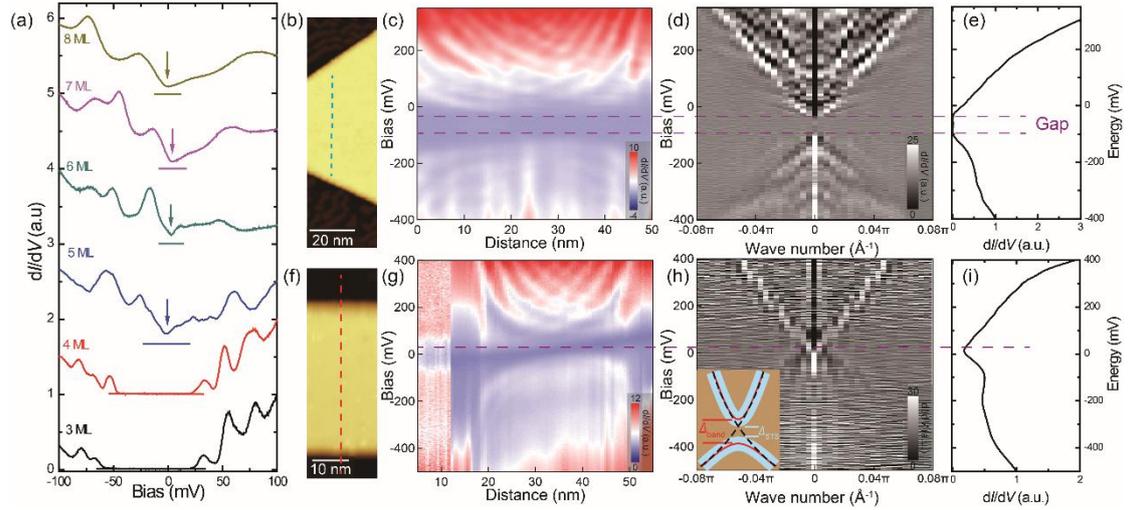

**Figure 2.** **(a)** A series of d$I$/d$V$ spectra recorded on Na$_3$Bi terraces with different thickness ranging from 3 MLs to 8 MLs. All spectra are shifted vertically for clarity. The short horizontal bars with the same color mark the zero differential conductance for each curve. **(b)** STM topography images of an isolated 3-ML Na$_3$Bi island, supported on a graphene substrate ($V_{bias}$ = +3.0 V, $I_t$ = 10 pA). **(c)** d$I$/d$V$ spectra taken along the cyan and red dashed lines in (b). **(d)** Energy dispersion relation of the standing-waves in (c) by 1D fast Fourier transformation. To obtain better resolution, second derivation is additionally performed to enhance low intensity features. **(e)** Average STS over the 3-ML Na$_3$Bi terrace. **(f)-(i)** The same as (b)-(e) but on an isolated 5-ML Na$_3$Bi island. The horizontal dashed purple lines indicate the gap edges. Inset in (h) is a sketch for band gap determination. Dashed black curves are the pristine Dirac cone for bulk Na$_3$Bi. Red lines indicate a band gap is opened near the Dirac node for Na$_3$Bi films ($\Delta_{band}$). Considering the effects of tunneling background, lifetime broadening and disorder, the bands become broadening and show a smaller gap in STS experiments ($\Delta_{STS}$).

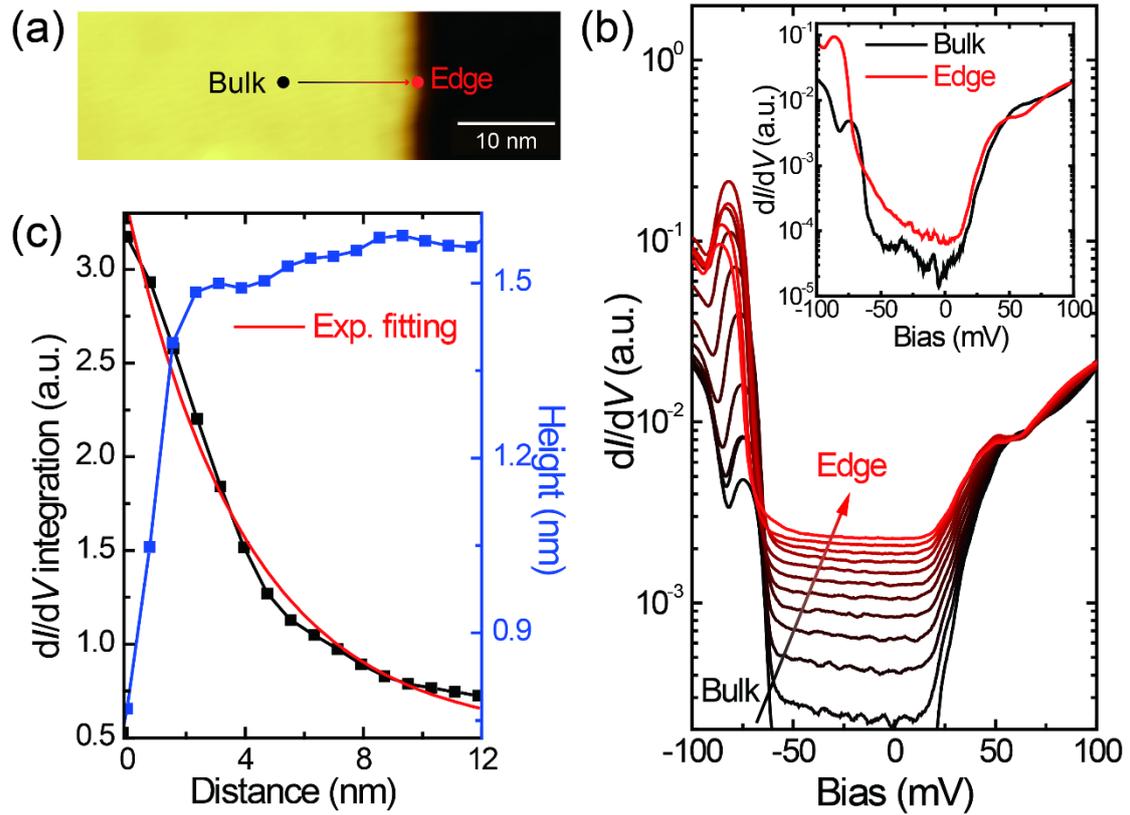

**Figure 3. (a)** STM image of a 3-ML Na$_3$Bi step on graphene ($V_{bias}$ = +3.0 V, $I_t$ = 10 pA). **(b)** A series of d$I$/d$V$ spectra (logarithmic scale) recorded near the step edge in (a). All spectra are shifted vertically for clarity. Inset is a direct comparison for the bulk spectra with one at the edge. **(c)** Intensity profile of integrated conductance within the gap (black) and topographic height across the step (blue). The edge state shows an exponential decrease into the bulk, with an exponential fitting of $y = 0.51 + 2.8*\exp(-x/4.06)$ (red line).

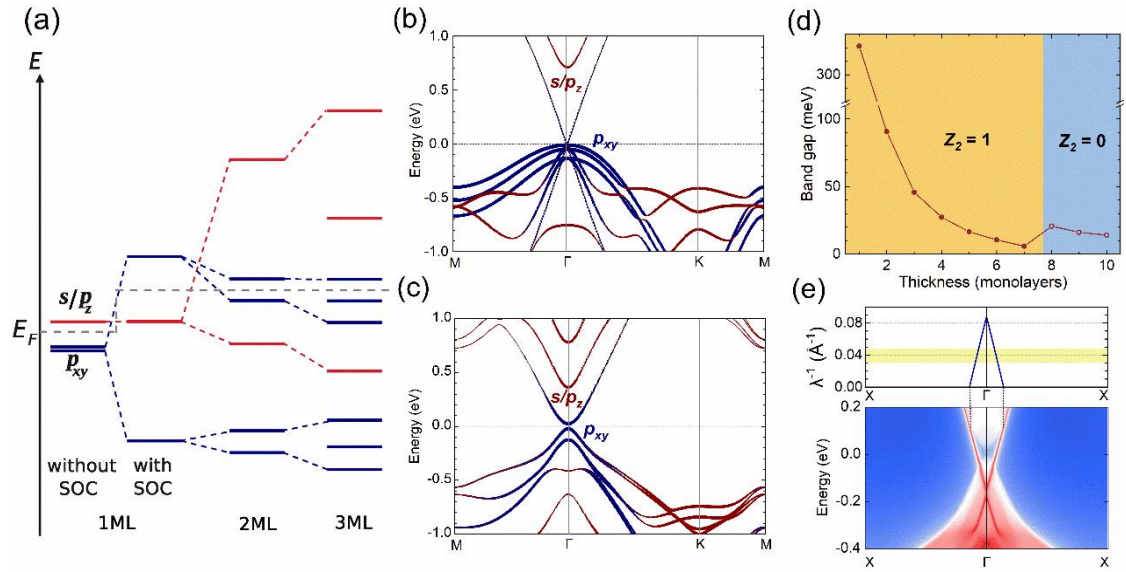

**Figure 4. (a)** Schematic diagram of orbital evolution at the Γ point for $Na_3Bi$ films from 1 ML to 3 MLs, which illustrates the effects of SOC and interlayer coupling. Calculated band structure of 3-ML $Na_3Bi$ **(b)** without and **(c)** with SOC. Contributions of the $s/p_z$ orbitals of Na and Bi atoms and the $p_{xy}$ orbitals of Bi are denoted by red and blue dots, respectively. **(d)** The calculated band gaps and $Z_2$ values as a function $Na_3Bi$ film thickness. **(e)** Topological edge states of 3-ML $Na_3Bi$ and their penetration depth (λ) as a function of momentum (the in-gap region is shaded yellow).